\documentclass[aps,prd,twocolumn,showpacs,superscriptaddress,groupedaddress]{revtex4}

\usepackage{color}
\usepackage{float}
\usepackage{graphicx}  
\usepackage{dcolumn}   
\usepackage{bm}        
\usepackage{amssymb}   

\newcommand{\sun}{\odot}

\begin{document}

\widetext
\leftline{Version 01 as of \today}
\leftline{Primary author: Lin F. Yang, Physics}
\leftline{Comment to {\tt lyang@pha.jhu.edu}}

\title{The Dark Matter Contribution to Galactic Diffuse Gamma Ray Emission }

\author{Lin F. Yang}
\affiliation{Department of  Physics \& Astronomy, The Johns Hopkins University \\
3400 N Charles Street, Baltimore, MD 21218, USA}

\author{Joseph Silk}
\affiliation{Department of  Physics \& Astronomy, The Johns Hopkins University \\
3400 N Charles Street, Baltimore, MD 21218, USA}
\affiliation{Institut d'Astrophysique de Paris - 98 bis boulevard Arago -
75014 Paris, France}
\affiliation{Beecroft Institute of Particle Astrophysics and Cosmology, Department of Physics,
University of Oxford, Denys Wilkinson Building, 1 Keble Road, Oxford OX1 3RH, UK}

\author{Alexander S. Szalay}
\affiliation{Department of  Physics \& Astronomy, The Johns Hopkins University \\
3400 N Charles Street, Baltimore, MD 21218, USA}

\author{Rosemary F.G. Wyse}
\affiliation{Department of  Physics \& Astronomy, The Johns Hopkins University \\
3400 N Charles Street, Baltimore, MD 21218, USA}

\author{Brandon Bozek}
\affiliation{Department of  Physics \& Astronomy, The Johns Hopkins University \\
3400 N Charles Street, Baltimore, MD 21218, USA}

\author{Piero Madau}
\affiliation{Department of Astronomy and Astrophysics, University of California, Santa Cruz, CA 950064, USA}

\begin{abstract}

Observations of diffuse Galactic gamma ray emission (DGE) by the
 {\it Fermi} Large Area Telescope (LAT) allow a detailed study of cosmic rays
and the interstellar medium. However, diffuse emission models of the
inner Galaxy underpredict the Fermi-LAT data at energies above a few GeV
and hint at possible non-astrophysical sources including dark matter
(DM) annihilations or decays. We present a study of the possible
emission components from DM using the high-resolution Via Lactea II
N-body simulation of a Milky Way-sized DM halo. We generate full-sky
maps of DM annihilation and decay signals that include modeling of
the adiabatic contraction of the host density profile, Sommerfeld-enhanced DM annihilations, $p$-wave annihilations, and decaying DM. 
We compare our results with the DGE models produced by the Fermi-LAT
team over different sky regions, including the Galactic center, high
Galactic latitudes, and the Galactic anti-center.  
This work provides possible {smooth component} templates of DM
to fit the observational data. 
{ The subhalo contributions can be considered to provide
statistically meaningful templates, and demonstrate how spatial profiles are significantly  modified according to  different annihilation/decay scenarios.
We argue that a  subhalo-based approach can help constrain the DM physics.
} 

\end{abstract}

\keywords{Dark Matter, Diffuse Gamma Ray}

\pacs{95.35.+d}

\maketitle

\section{Introduction}
Overwhelming observational evidence \citep{Frenk:2012dq} 
implies the existence of dark matter (DM), whose nature is still
unknown. Searching for DM is now a major research theme in both
particle physics and astrophysics. The former mainly focusses on
direct detection by experiments sensitive to the interaction of
DM particles with normal matter, while the latter uses indirect
methods, including measurements of the end products of DM
annihilation/decay (e.g. $\gamma$-ray, $e^+e^-$ pairs) or mapping of
gravitationally lensed structures.  A number of particle physics
scenarios have been proposed for the nature of DM. Among these
scenarios, the most compelling is that of the weakly interacting massive
particle (WIMP) that acts as cold dark matter and effectively explains
the origin of the large-scale structure of the universe.  Theories generally 
hypothesize  WIMPs that interact with normal matter no more
strongly than the weak force and therefore direct observation is
challenging.

If DM is a form of a thermal relic
particle that was once in thermal equilibrium in the very early
universe, then it may light up Galactic substructures by
self-pair annihilations.  
{
In addition, DM decay 
products may also be a source that is comparable to
annihilation.
}  
The absolute $\gamma$-ray intensity depends on the WIMP
annihilation cross section or decay rate, particle type,  particle mass, and
astrophysical distributions, all of which are poorly known. The
relative annihilation/decay luminosity is density-dependent, 
which may offer insight into the nature of DM, e.g. 
the decay signal is proportional to density while the annihilation luminosity
is proportional to density squared. Additional correction factors may be applied to
the standard velocity-weighted thermal cross section $\langle\sigma
v\rangle$. One example is the Sommerfeld boost -- a nonperturbative increase in the annihilation cross-section at low 
velocities that is the result of a generic attractive force between the incident DM particles. 
Baryons could also play an important role in determining the profiles of DM halos. Adiabatic contraction
induced by the infall of baryonic matter could steepen the DM density profile
substantially, e.g. to $r^{-1.9}$ from a Navarro-Frenk-White (NFW) \citep{Navarro:1996ce}  profile $r^{-1}$
\citep{Freese:2009ic}. Another possible correction to the annihilation
cross-section that has not been previously discussed in the galactic
context is that of $s$-wave suppressed annihilations, as might be
important for spin-dependent interactions.

Diffuse galactic $\gamma$-ray emission (DGE) is believed to be
produced by interactions between cosmic rays and the interstellar
medium (ISM). The launch of the Fermi Gamma-ray Satellite, with its Large Area Telescope (LAT),
enabled a detailed study of cosmic-ray origin and propagation, and of
the interstellar medium. In a recent study \citep{Ackermann:2012co},
the Fermi-LAT team published an analysis of the measurements of the diffuse
$\gamma$-ray emission from the first 21 months of the Fermi
mission. They compared the data with models generated by the
GALPROP\footnote{http://galprop.stanford.edu/} code, and showed that these emission templates
underpredict the Fermi-LAT data at energies of a few GeV in the inner galaxy.  
This can possibly be explained by undetected point-source populations and 
variations of the cosmic ray spectra. \citet{Ackermann:2012ha}
have performed an analysis of the DGE in the Milky Way halo region to search for a
signal from dark matter annihilations or decays. They considered a large 
region covering the central part of the Galactic halo while masking 
out the Galactic plane. In such a region, the DM signal would have a large S/N
ratio and would not depend on detailed  assumptions about the center profile, e.g. 
the assumptions of  a NFW profile or a  cored profile would only differ by a 
factor of $\sim 2$. This paper provided conservative limits on the DM cross-section
assuming that all the $\gamma$-ray signal comes from DM in this region.
They also provided more stringent limits based on modeling the 
foreground $\gamma$-ray signal with the GALPROP code. Their results
impact the possible mass range over which  DM is produced thermally in the early universe and
challenge the DM annihilation interpretation of PAMELA/Fermi-LAT cosmic ray 
anomalies. However, several points could be improved: 1) although using a NFW profile for modeling pure DM  and a 
cored profile for modeling the baryonic matter effect covers extreme cases, it is still worth trying a more 
realistic profile directly from simulations;  2) in addition to the mass and cross-section of the DM particle, 
the different annihilation/decay schemes may also play an important role, e.g. subhalo effects would become considerably 
strong in the region of interest where tidal disruption plays a role and especially   when considering possible Sommerfeld 
enhancements;  and 3) instead of detecting DM signal from the region that masks out the Galactic Center region, it may be of 
interest  to search for a possible signal from the center or anticenter  regions. Therefore, a good template for the Galactic 
Center or the full sky should provide more constraints on the DM physics; 4) the spatially dependent velocity dispersion of 
the DM due to substructure plays an important role both in considering Sommerfeld or $p-$wave enhanced  annihilation signals, 
and simulations can provide this information.

The contribution of DM annihilation/decay processes to the Fermi-LAT $\gamma$-ray signal in the Galactic Center (or its expected 
signal from subhalos in the Galactic halo) is difficult to quantify without knowing the galactic distribution of dark matter 
and the dark matter particle properties leading to annihilations or decays. To compare the observations with predictions of DM 
emission, we must rely on numerical simulations that follow the formation of cosmic structure in the highly non-linear regime. 
A few high-resolution simulations of Milky Way-sized halos have been completed over the last few years: the Aquarius project \citep{Springel:2008gd}, 
the Via Lactea series \citep{Diemand:2008hr}, and the GHALO run \citep{Stadel:2009el}. The ability of the Fermi-LAT satellite to detect DM annihilation 
signals from  Galactic subhalos has been previously studied by \citet{Anderson:2010ds} and \citet{Kuhlen:2008kr} using the high-resolution Via Lactea 
II (VLII) N-body simulation. To date, there has been no clear signal of dark matter annihilations or decays in the outer galactic halo. There are numerous possible 
explanations for non-detection in the outer galactic halo combined with the observed DGE in the Galactic Center, including simulated DM-only structure 
being inconsistent with the actual galactic structure, the predicted DM annihilation/decay luminosity from particle physics being overly optimistic, or 
some misunderstanding of the observational data. 
{
Motivated by the range of possible explanations for the excess of DGE data to the model, we present a  study of various possible Galactic DM contributions to the observed DGE emission. 
}
We construct several maps of gamma-ray emission from different DM 
annihilation/decay scenarios in order to illustrate  how each scenario may offer additional templates to fit the Fermi data. We additionally compared 
DM annihilation signals in each scenario to the DGE model from the GALPROP code of \citet{Ackermann:2012co} in different regions on the sky.

We organize the rest of this paper as follows: in $\S$\ref{sec:darkmatter}, we give a
description of different DM annihilation/decay schemes.  Our data
processing method and $\gamma$-ray generating algorithm are described
in $\S$\ref{sec:simulation}.  We show how we add our correction to the normal
annihilations/decays in $\S$\ref{sec:corrections}. The all-sky maps results and 
comparisons to Fermi data for different sky regions are presented in \S\ref{sec:regions}. 
Conclusions and discussion are presented in \S\ref{sec:discussion}.

\section{DARK MATTER ANNIHILATION AND DECAY\label{sec:darkmatter}}

 \begin{figure*}
    \centering
     \includegraphics[width=0.95\textwidth]{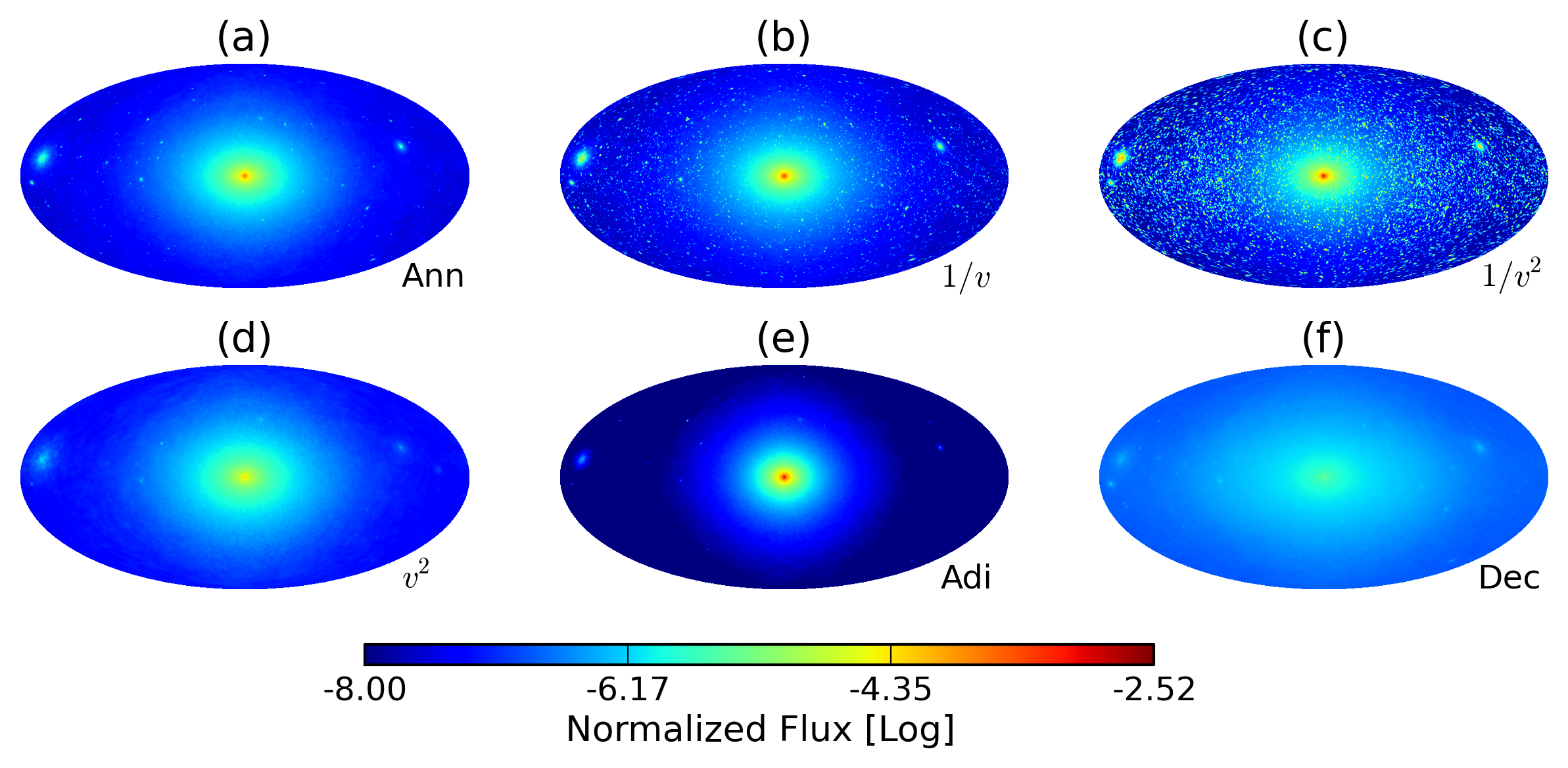}
    \caption{Mollweide projection of DM annihilation flux. {\it From (a) to
    (f)}: pure annihilation, annihilation with Sommerfeld enhancement
    $1/v$ correction, annihilation with Sommerfeld enhancement $1/v^2$
    correction, annihilation with $v^2$ correction, annihilation with
    adiabatic contraction, pure decay. A Gaussian filter with FWHM
    $0.5^\circ$ is applied. The flux is normalized such that the host
    halo has flux unity.}
    \label{fig:allskymap}
    \end{figure*}
 
Gamma-rays are one of the final products from DM self-annihilations
or decays.  In the case of annihilations, the $\gamma$-ray flux (
photons cm$^{-2}$ sr $^{-1}$s$^{-1}$ ) in a solid angle d$\Omega$ of a given
line-of-sight $(\theta,\phi)$ can be written as
\begin{equation}
	\Psi_\gamma(\theta, \phi)\propto {\langle\sigma v\rangle}\int_{los}{\rho{(\bm{r})}^2dr},
	\label{eqn:dm_ann}
\end{equation}
where  $\langle\sigma
v\rangle$ is the velocity-weighted thermal cross  section, which is
usually treated as a constant, and $\rho(\bm{r})$ is the density of DM
at position $\bm{r} = {r(\sin\theta\cos\phi, \sin\theta\sin\phi,
\cos\theta)}$, and $r$ is the distance from the DM particle to the
observer.  In the case of decay, the $\gamma$-ray intensity is then
proportional to $\rho$,
\begin{equation}
	\Psi_\gamma(\theta, \phi) \propto \frac{1}{ \tau} \int_{los}{\rho{(\bm{r})}dr}
	\label{eqn:dm_decay}, 
\end{equation}
 where $\tau$ is the decay lifetime.  Eqn-(\ref{eqn:dm_ann}) and Eqn-(\ref{eqn:dm_decay})
 are usually written as a combination of a particle physics factor $P$  and an astrophysical factor $J$, while 
 we combined them as $\Psi$ to indicate what has been considered in our $\gamma$-ray producing
 code. Since $\langle\sigma v\rangle$
 and $\tau$ are unknown, the normalization is arbitrary.  For a given
 astrophysical DM distribution, the $\Psi$ factors given here
determine the relative intensity of the predicted flux.  Calculating
 the absolute value of $\Psi$ requires the absolute value of the
 particle physics parameters. As shown in several papers
 \citep{Berlin:2013vk, Feng:2010ck}, $\langle\sigma v\rangle$ is less
 than $\sim 10^{-26}$ cm$^2$s$^{-1}$, and particle mass $M_\chi$ varies from 100 GeV to 10
 TeV for the case of WIMPs. For our purpose, only the relative
 intensity is needed. The $\gamma$-ray maps 
 in this work are produced from the simulation data (see \S\ref{sec:simulation}) by using one of these formulae in addition to possible cross-section correction factors. We consider in
 total six different $\gamma$-ray map production scenarios: (a) pure
 annihilation without any 
 correction factors, assuming $\langle\sigma
 v\rangle$ is a constant; (b) annihilation with a Sommerfeld enhancement
  of $1/v$ included in the annihilation cross section ($v$ is the
 relative velocity of the annihilating particles); (c) annihilation with
a Sommerfeld enhancement of $1/v^2$, as expected near resonance; (d)
annihilation proportional to $v^2$ consistent with $s-$wave channel suppression ; 
(e) adiabatic contraction to the dark matter density profiles, and 
(f) pure decay, taken to be linear in dark matter density.
The correction cases are described in $\S$\ref{sec:corrections}. 
 
\section{SIMULATION DATA AND MAP PRODUCTION ALGORITHM \label{sec:simulation}}

The Via Lactea series includes some of the highest resolution collisionless simulations  of the assembly of a Milky Way-sized halo. 
VLI, the first of the series, contains $2\times 10^8$ DM particles, covering the virial volume and surroundings
of a host halo of $M_{200} = 1.77\times10^{12}\rm{M}_\sun$. The host halo and subhalo properties are presented 
in \citet{Diemand:2007fi}, \citet{Diemand:2007hb}, and \citet{Kuhlen:2007bf}. For this work, we used the second 
generation of the simulation series, the VLII run \citep{Diemand:2008hr, Kuhlen:2008kr}, which has a slightly higher
resolution. It employs about 1 billion particles each of mass $4,100\rm{M}_\sun$ to simulate a Milky-Way-sized host 
halo and its substructures.  We extract the  roughly $4\times10^8$ particles of the $z=0$ snapshot within $r_{200}$ ($402$ kpc,
the radius where the density is 200 times larger than the critical density $\rho_c$) of the host halo. About $5\times 10^4$ 
individual subhalos have been identified in this region. The halos found within $r_{200}$ show considerable self-similarity. {
An 
NFW-like profile fitting to the density could be generalized by}, 
\begin{equation}
\label{eqn:nfw}
\rho(r) = \frac{\rho_s}{\left(\frac{r}{r_s}\right)^{\gamma}\left[1+\left(\frac{r}{r_s}\right)^\alpha\right]^{\frac{\beta-\alpha}{\alpha}}},
\end{equation} 
{
\citet{Diemand:2008hr}, fixes $\alpha=1$,   $\beta = 4 - \gamma$ and gives the best-fitting parameters $\gamma = 1.24$, $r_s = 28.1$ kpc and 
$\rho_s = 3.50\times10^{-3}\mathrm{M}_\sun$pc$^{-3}$}.  The host halo
is not spherically symmetric but is ellipsoidally shaped. The detailed host halo properties can be found in \citet{Diemand:2008hr}. 

To produce the $\gamma$-ray maps, we follow \citet{Kuhlen:2008kr} and locate a fiducial `observer' 8 kpc
from the host halo center along the intermediate principle axis of the ellipsoid. Given the position 
of an observer, the density field of each particle can be represented by a Dirac-$\delta$ function.
Eqn-(\ref{eqn:dm_ann}) and (\ref{eqn:dm_decay}) can be rewritten as the summation of flux over each 
particle, i.e. $\sum_iF_i$. For annihilation, $F_i\equiv{m_i\rho_i/4\pi r_i^2}$; and decay, 
$F_i\equiv{m_i/4\pi r_i^2}$, where $r_i$ is the distance from the particle to the observer. 
To avoid shot noise, we smoothed the particles with a SPH kernel so that each particle is represented by a
sphere of radius $h_i$ instead of a point. The flux $F_i$, correspondingly, spreads out as a Gaussian on 
the sky. We computed $\rho_i$, $h_i$ and $\sigma_i$ (the velocity dispersion) by finding the volume 
encompassing the nearest 32 particles (using the code SMOOTH).  The velocity dispersion $\sigma_i$ 
is read out simultaneously from this process.  

{Our  final result for the omnidirectional $\gamma$-ray signal  has been calculated using  HEALPIX 
\citep{Gorski:2005kua}. We use a NSIDE=512 to model the angular resolution of Fermi-LAT (roughly 0.2$^\circ$). 
This is an extremely compute-intensive task, it takes more than 8 hours on a regular CPU to produce a single map with the 400M  particles. In order to be able to create the maps involving the different annihilation scenarios and different viewpoints,  we have developed a novel GPU-based algorithm\citep{Yang:2013wf}. 

In this method, we projected $F_i(\theta, \phi)$  onto two separate tangential planes of the two celestial hemispheres using stereographic projection, then remapped these to the final HEALPIX projections. We have realized that the problem can be reduced to rendering a projected density profile of a particle, weighted by different factors depending on the physics of the annihilation. As these computations became quite similar to those used in visualization,  we were able to implement them in the shader language of OpenGL (Open Graphics Library) on a high-end Nvidia GPU for maximum performance. As a result, we are able to render the projected profiles at a rate of more than 10M particles per second and hence build a map in 24 seconds. This thousand-fold speedup was an essential factor, that enabled us to carry out the large number of numerical experiments needed to complete this paper. }

Even with the high resolution of VLII, only a portion of
the hierarchal structure of DM clumps is resolvable. In principle, the WIMP DM
substructure has mass power all the way down to the mass of the DM
kinetic or thermal decoupling scales, of order $\sim 10^{-6}\rm
M_\odot.$ These unresolved structures will boost the final brightness
of the $\gamma$-ray emission according to the $\langle\sigma v\rangle\rho^2$ 
dependence {
(while decay is not affected due to the $\rho$ dependence)}. Detailed studies of the boost factor $B(M)$ can be found in \citet{ Kuhlen:2008kr, Anderson:2010ds}. 
 {
 Although as pointed out by   Springel et al \citep{Springel:2008ge} and Fornasa et al. \citep{Fornasa:2013ig}  the unresolved subhalo component provides a important source of emission, Springel et al also note that this unresolved smooth component is quite flat (i.e. the contrast of the brightest and faintest points is only  about 1.5). In practice, this component is unlikely to be  distinguishable from other emission components. To include the emission of unresolved subhalos, we suggest that use of an all-sky uniform component would be an alternative solution.  On the other hand, the boost factor is also highly model-dependent (e.g. Kamionkoski et al. \citep{Kamionkowski:2010dy}). In terms of simplicity and clarity, we choose to not include the boost factor calculation in this work, since we only care about the possible shapes of the profiles.}
\begin{figure}
    \centering
    \includegraphics[width=0.49\textwidth]{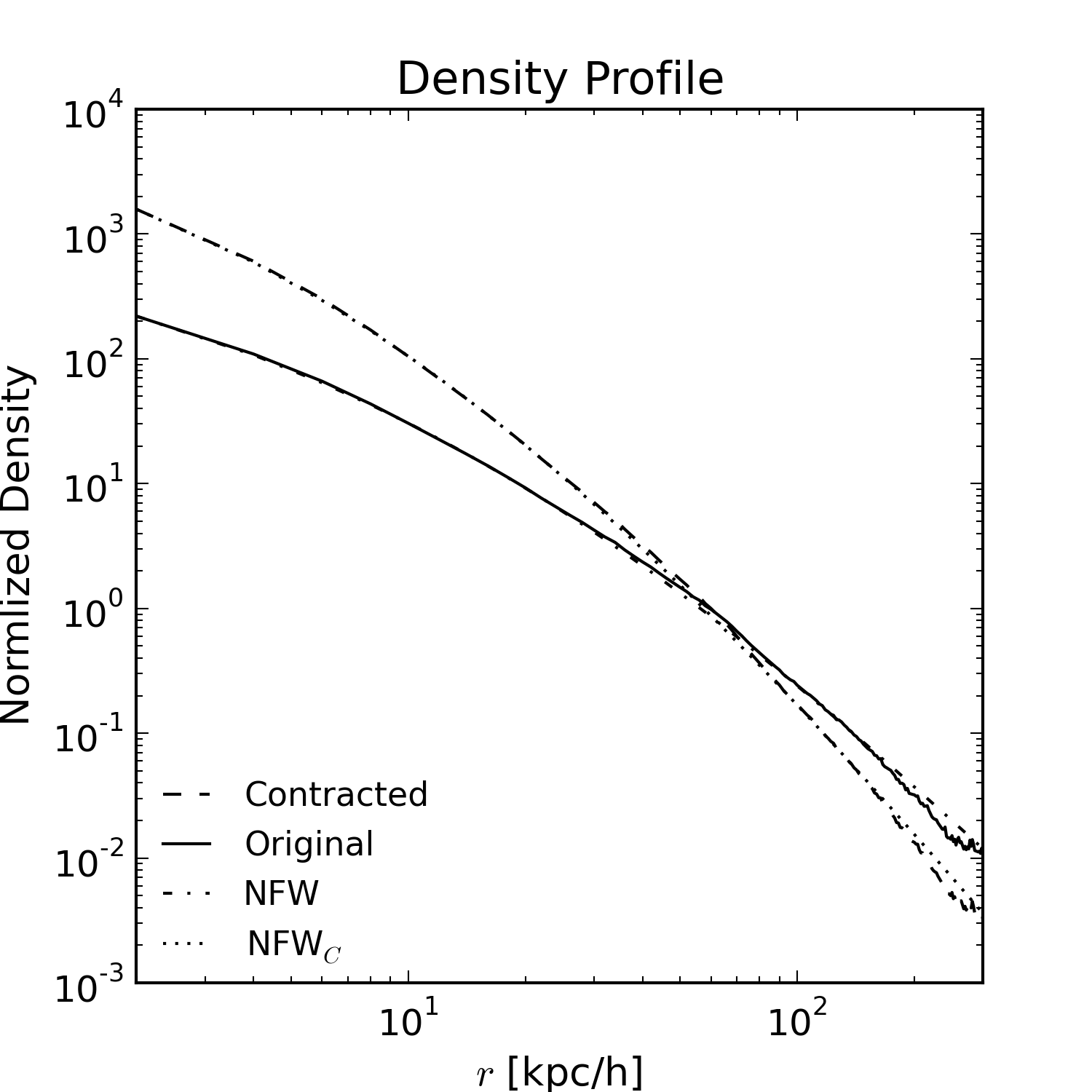}
    \caption{The density profile of the host halo and contracted
    profile. The solid black line is the original density profile of
    the host halo. After contraction, the profile becomes the long
    dashed line. The dotted line is the NFW profile with parameters
    given by \citet{Diemand:2008hr}, with $\gamma = 1.24, \alpha=1$
    and $\beta=1.76$. The dotted dash line is the NFW$_C$ profile
    describing the adiabatic contraction with $\gamma=1.37,
    \alpha=0.76$ and $\beta=3.3$. The normalization is chosen such
    that the intersection point of the contracted and non-contracted profiles are at a density of
    1. See text for details.}
    \label{fig:densityprofile}
\end{figure}

\begin{figure*}
    \centering
    \includegraphics[width=0.95\textwidth]{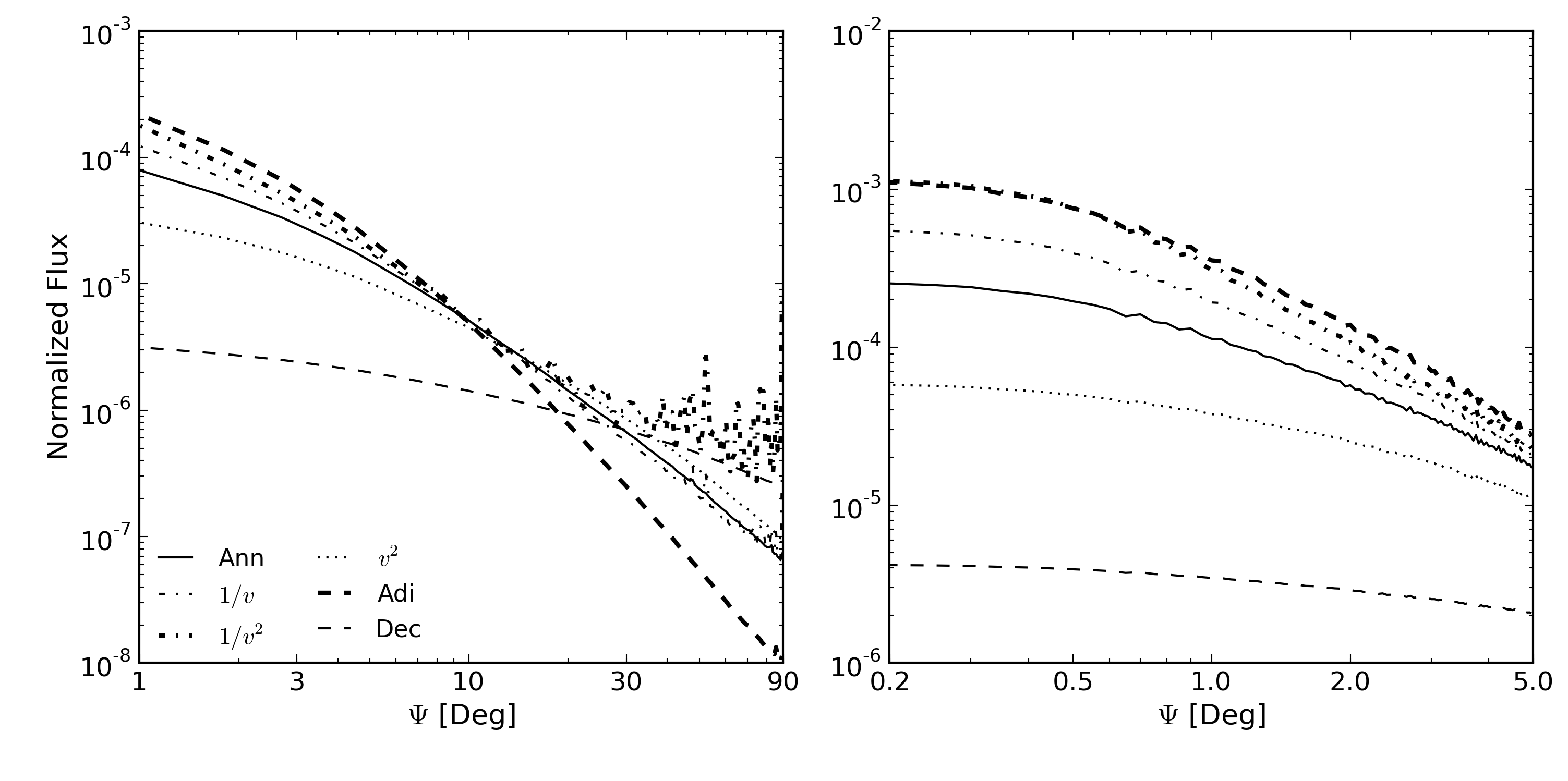}
    \caption{The flux annular profile of different cases. {\it Left panel}:
    the angular profile of $0-90^\circ$ region of the sky maps; {\it right panel}: the profile of
    the zoomed in region of the left panel in the inner
    $0-5^\circ$. For legends of lines: solid, pure annihilation; thin
    dash dotted, annihilation with Sommerfeld enhancement $1/v$
    correction; thick dash dotted, annihilation with Sommerfeld
    enhancement $1/v^2$ correction; dotted, annihilation with $v^2$
    correction; thick dashed, annihilation with adiabatic contraction;
    thin dashed, pure decay. The flux are normalized such that the
    host halo has flux unity. See text for details. }
    \label{fig:fluxprof}
\end{figure*}

\section{CORRECTIONS\label{sec:corrections}}
\subsection{Sommerfeld Enhancement}

The PAMELA \citep{Adriani:2009eh} and AMS-02 results
\citep{Aguilar:2013hm} show a rise in positron fraction at high
energy and a hardening of the spectral index, while no antiproton excesses
were found. If this signal were due to annihilation, the standard
thermal cross-section $\langle\sigma v\rangle$ would be too small to
simultaneously fit $e^+e^-$ and avoid antiproton excesses. To gain a
larger cross-section, a force carrier $\phi$ \citep{Feng:2010ck,
Lattanzi:2009dg, ArkaniHamed:2009gk} is proposed for mediating the DM
particle interaction. This mediator could be a standard model particle
or an unknown boson responsible for dark sector forces. Assuming the
DM particle is a Majorana or Dirac fermion, such a process is denoted
as $XX\rightarrow\phi\phi$. The Sommerfeld enhancement
\citep{Sommerfeld:1934} can be therefore calculated. For the resonance
case (when $M_X/m_\phi \approx n^2/\alpha $, where $\alpha$ is the
coupling strength, and $n$ is  an integer), this model gives an
enhancement  of $S=1/v^2$ to the cross-section. For the non-resonance
case, $S\approx 1/v$.  For $c/v \approx 1$, $S\approx 1$.  The
Sommerfeld cross-section must saturate in a viable model: we take this
saturation to be $\sim 1$ km/s, above the resolution limit of our
simulations. From our point of view, this amounts to a renormalisation
that we ignore: it is the profile shape and clumpiness that we care
about.  The cross-section for DM particle pair annihilation changes
significantly over the simulation range caused by velocity changes in different environments,
 i.e. by a factor $\sim10^2$ or more \citep{Kuhlen:2009cn}. In such a case,
  the subhalos, whose velocity dispersion is less than that of the central
halo, light up. For the $1/v^2$ case, the substructures almost
dominate,  as shown in Fig-\ref{fig:allskymap}(c). 

To calculate an accurate Sommerfeld enhancement value at a given
location, one needs the whole phase-space distribution of each
particle. This is not possible for our case using the VLII simulation. We
therefore follow \citet{Kuhlen:2009cn} and assume a Boltzmann velocity
distribution of the relative velocity $v_\mathrm{rel}$:
\begin{equation}
f(v_\mathrm{rel}, \sqrt{ 2/3}\sigma_v) = 4\pi/(2\sqrt{2/3}\pi\sigma_v)^{3/2}v_\mathrm{rel}^2
\exp[-3v_\mathrm{rel}^2/\sigma_v^2].
\end{equation}
The Sommerfeld enhancement is then given by $S(\sigma_v) = \int{f(v, \sqrt{2/3}\sigma_v) S(v)dv}$. 
If $S(v)$ has no saturation, then $S(\sigma_v)$ is then $\sim 1/\sigma_v$ for the $1/v$
case and $1/\sigma_v^2$ for the $1/v^2$ case. If one takes into
consideration the saturation velocity $v_s$, below which $S(v)$ no
longer increases, we find $S(\sigma_v)$ presents similar properties. We
therefore approximate $S(\sigma_v)$ completely by $S(v)$ without losing
much accuracy. In our calculation, we calculated $S$ as $1/\sigma_v$ or
$1/\sigma_v^2$ when $\sigma_v>v_s$, and as $S(v_s)$ when $\sigma_v < v_s$.
Here $\sigma_v$ is the velocity dispersion mentioned in the last
section. Fig-\ref{fig:allskymap} (b) and (c) shows the Sommerfeld
enhancement case for the $\gamma$-ray map. To avoid the cross-section blowing up at very
low velocity, we applied saturation velocities equal to  1 km/s for the $1/v$
case and 5 km/s for  the $1/v^2$ case.

\subsection{p-wave annhiliation}

As indicated in the Sommerfeld enhancement case, the relative velocity
of particles varies significantly after thermal freeze out. If the
cross-section is velocity-dependent, it will either suppress high
velocity or low velocity annihilation. Constraints on annihilation processes from CMB and
$\gamma$-ray observations could be considerably weakened if the $s-$wave channel were suppressed. $p-$wave annihilations are orders of magnitudes
larger than $s-$wave annihilations at recombination and are an interesting
case to be considered. For example, neutralino annihilation could be dominated by the $p-$wave process \citep{Goldberg:1983gk}. In contrast to the Sommerfeld case, we consider an annihilation case whose $s-$wave process is suppressed. For the annihilation rate,
we phenomenologically set $\langle\sigma v\rangle \propto
v^{2(n-1)}$, where $n=1(2)$ for $s-$wave($p-$wave) annihilation
\citep{Essig:2013td}. Fig-\ref{fig:allskymap} (d) shows a map from
the $p-$wave annihilation with the $s-$wave channel suppressed. More discussion will
be given in \S\ref{sec:discussion}.

\subsection{Baryonic matter and adiabatic  contraction}

When structure is forming, baryonic matter dissipates its thermal and
kinetic energy and falls  to the center of the DM halo. The
condensation of gas and stars in the inner regions of DM halos will
adiabatically contract the DM density distribution and lead to a
denser profile in the center. This effect is implied from both
theoretical and observational studies \citep{Gnedin:2011tf}.  Adiabatic contraction is generally studied using the
standard contraction model (SAC) as introduced by
\citet{Blumenthal:1986ie}. However, hydrodynamic
simulations show that the contraction is weaker and a modified adiabatic
contraction model was introduced in \citet{Gnedin:2004hc}. We  consider the
baryonic matter contraction effect only in the host halo, where there
is enough baryonic matter to cause the contraction. To simplify the
calculation, we follow \citet{GomezVargas:2013fj} and modify the host
halo density profile to be a NFW$_c$ profile with $\alpha = 0.76$,
$\beta = 3.3$, $\gamma = 1.37$ and $r_s = 18.5$kpc. To conserve the
the total mass of the host halo within $r_{200}$, we use
$\rho_s^\prime = 16.83\rho_{s}$. We convolve a contraction factor
$c(r) = \mathrm{NFW}_c(r)/\mathrm{NFW}(r)$ with the density of each
particle in the simulation. The mass of each particle is changed
accordingly. The contracted density profile and the
original profile are shown in Fig-\ref{fig:densityprofile}.  We
normalized the profile such that the intersection point of the
original and the contracted profile is at a density of $1$. We include the NFW and NFW$_c$ profile in Fig-\ref{fig:densityprofile} as  points of comparison.  
Fig-\ref{fig:allskymap} (e) shows
an annihilation map of the contracted host halo plus uncontracted
subhalos.

\begin{figure*}
    \centering
    \includegraphics[width=0.49\textwidth]{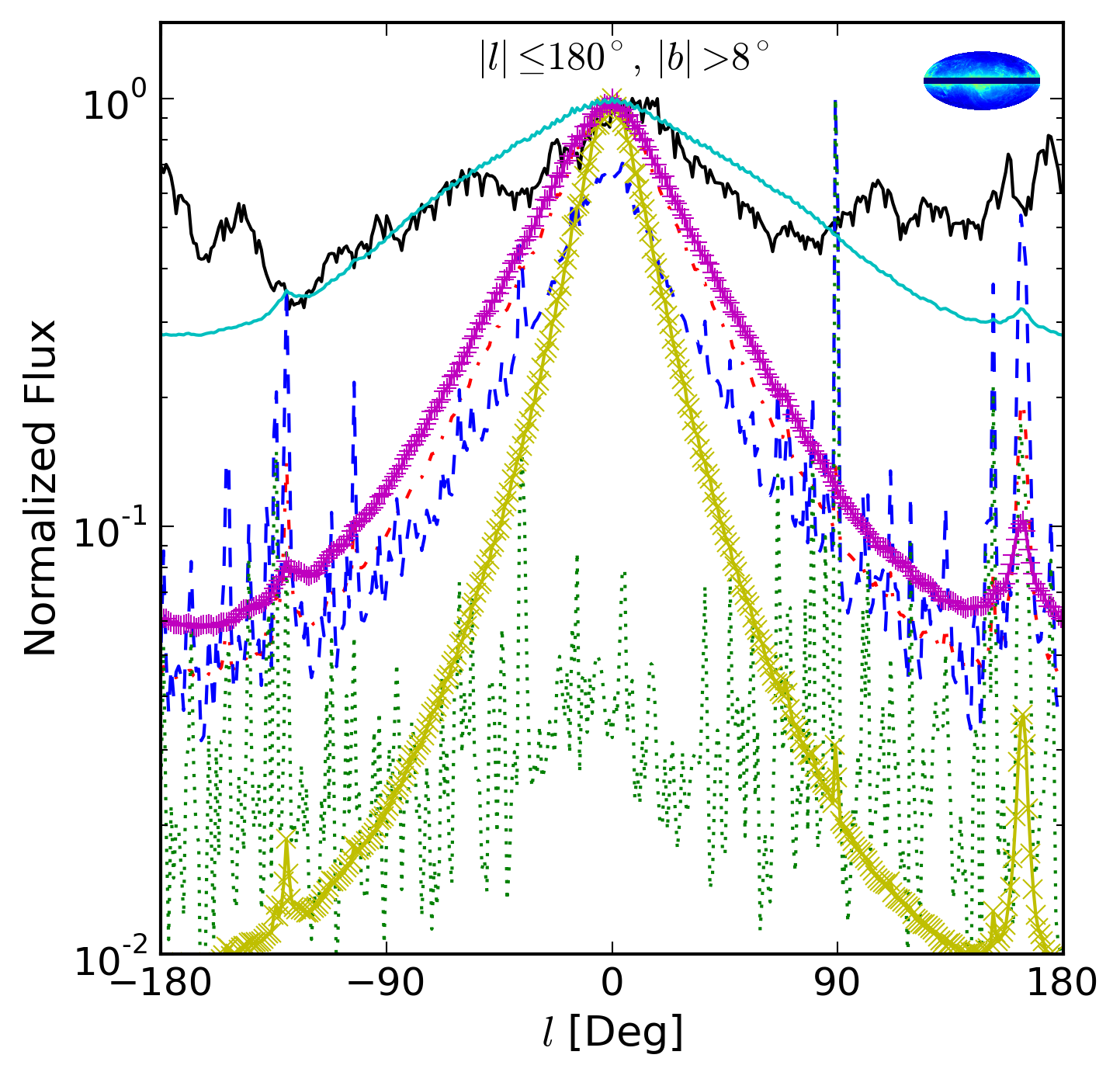}
    \includegraphics[width=0.49\textwidth]{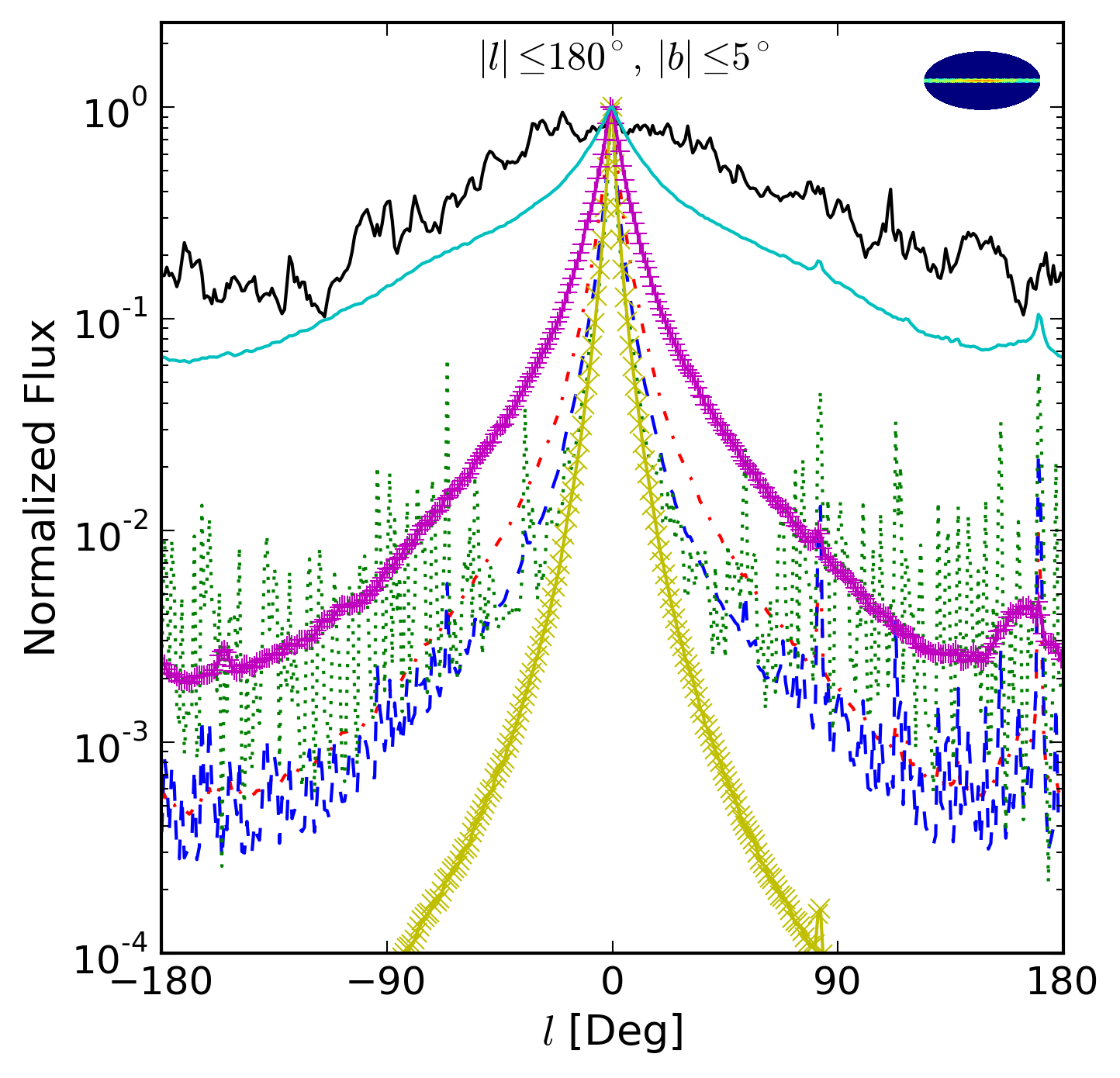}
    \includegraphics[width=0.49\textwidth]{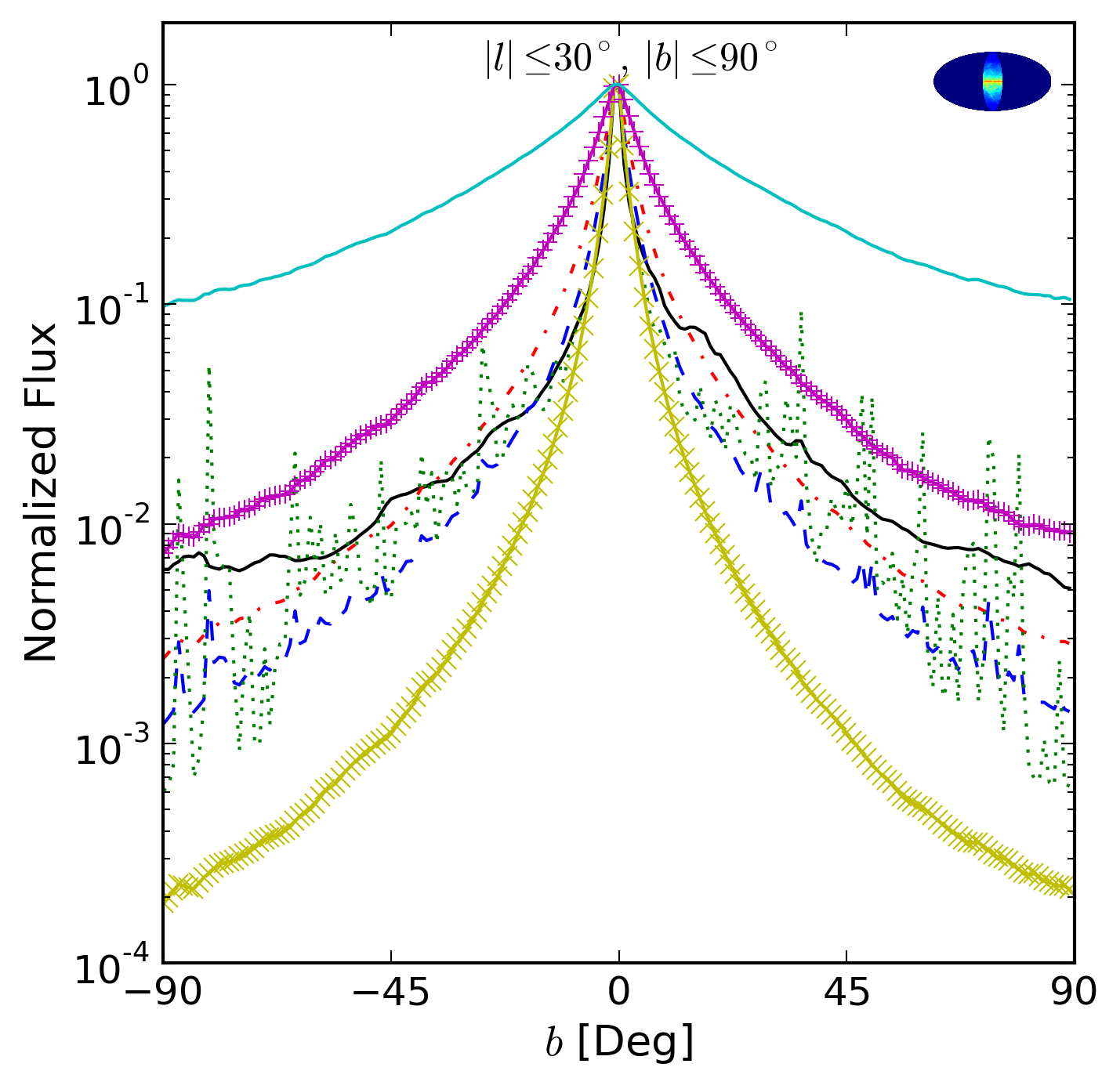}
    \includegraphics[width=0.49\textwidth]{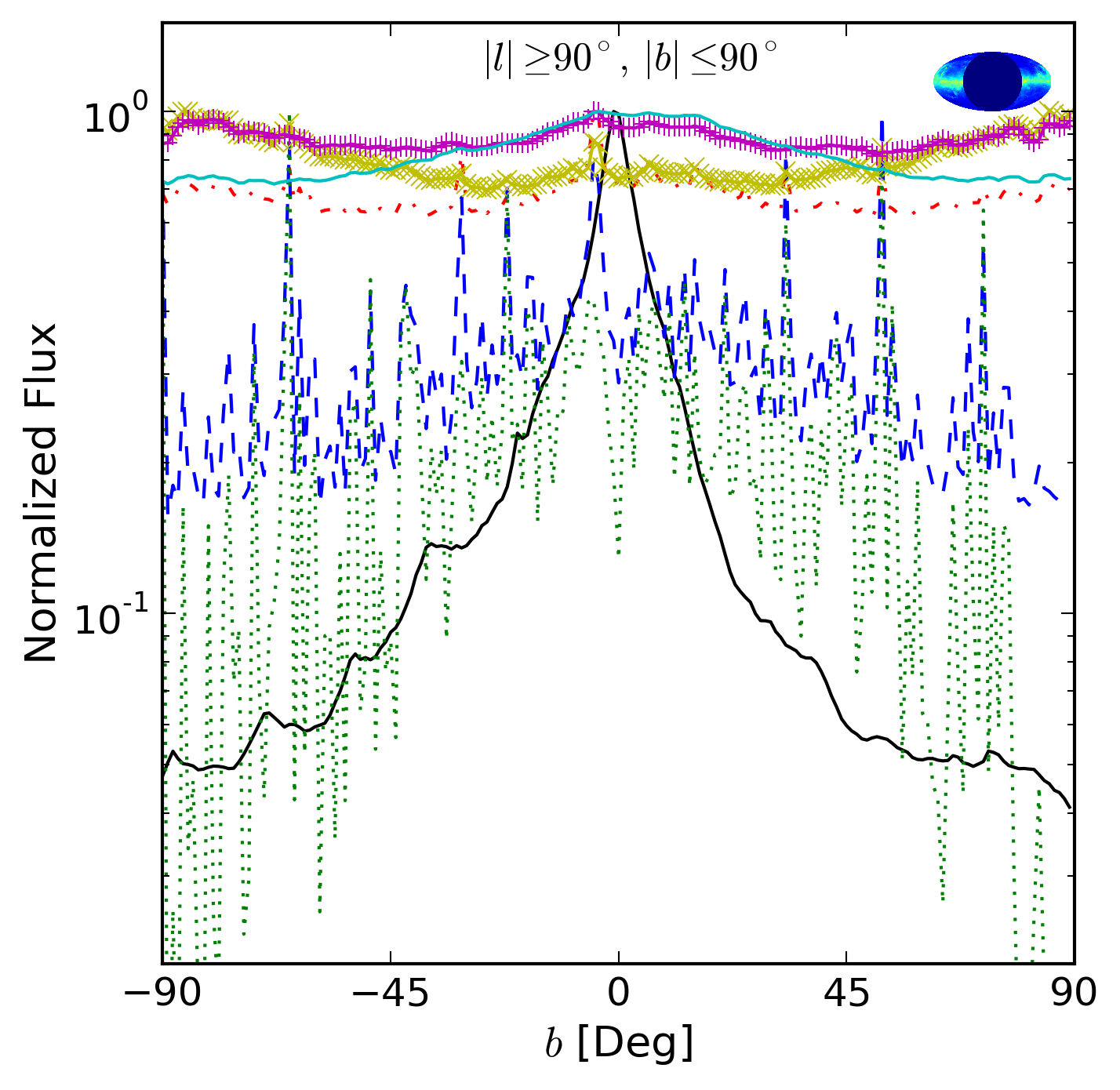}
    \caption{Angular profile of different sky regions. {\it Upper left panel}:
    the longitudinal profile of $|b|>8^\circ$; {\it upper right panel}: the
    longitudinal profile of $|b|\le5^\circ$; {\it lower left panel}:
    latitude profile of $|l|\le30^\circ$ and {\it lower right panel}:
    latitude profile of $|l|\ge90^\circ$, {
    this profile is particularly processed such 
    that there is no subhalo too close to the observer.} For legends of lines: black
    solid, the DGE model
    $^\mathrm{S}\mathrm{S}^\mathrm{Z}4^\mathrm{R}20^\mathrm{T}150^\mathrm{C}5$
    from Fermi-data; red dash dotted, pure annihilation; blue dashed,
    annihilation with $1/v$ correction; green dotted, annihilation
    with $1/v^2$ correction; yellow solid with ``x" symbols, annihilation
    with adiabatic contraction; magenta solid with ``+" symbols,
    annihilation with $v^2$ correction; cyan solid, pure decay. All
    profiles are normalized so that they have maximum unity. See text
    for details.}
    \label{fig:cutplots}
\end{figure*}

\begin{figure*}
    \centering
     \includegraphics[width=0.95\textwidth]{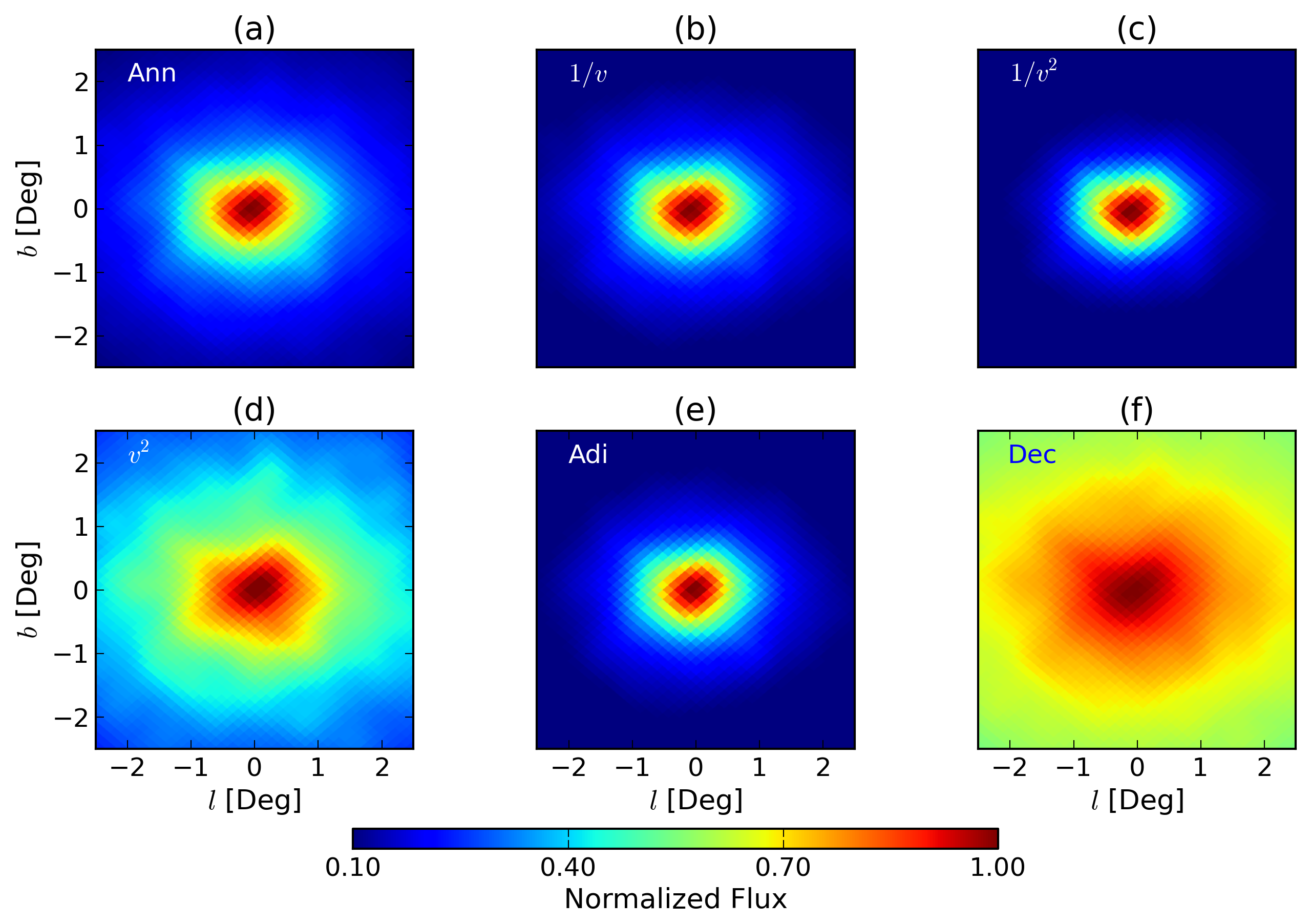}
    \caption{Color map of the $\gamma$-ray flux in the inner
    $5^\circ\times5^\circ$ produced by DM annihilation/decay.  From
    (a) to (f): pure annihilation, annihilation with Sommerfeld
    enhancement $1/v$ correction, annihilation with Sommerfeld
    enhancement $1/v^2$ correction, annihilation with $v^2$
    correction, annihilation with adiabatic contraction, pure
    decay. The flux are normalized such that the center pixel has flux
    unity.}
    \label{fig:center55}
\end{figure*}

\section{ALL-SKY MAPS AND GAMMA-RAY SIGNALS  FROM  DIFFERENT SKY REGIONS\label{sec:regions}}

The all-sky maps of the six different annihilation/decay scenarios are shown in Fig-\ref{fig:allskymap}.
The normalization is chosen such that the host halo has flux unity. 
All the maps are smoothed by a Gaussian kernel with FWHM 0.5$^\circ$. 
They show significant differences in both the center halo profile and the 
relative subhalo flux. Compared to the pure annihilation case of Fig-\ref{fig:allskymap}a,  
the $1/v$ and $1/v^2$ Sommerfeld enhancement cases (Fig-\ref{fig:allskymap}b) have considerably more substructures visible. 
The $1/v^2$ resonant Sommerfeld enhancement case has the largest contribution from substructures to the annihilation flux 
of the six scenarios. For the $p-$wave annihilation case ( Fig-\ref{fig:allskymap}d),  the adiabatic contraction case (Fig-\ref{fig:allskymap}e) 
and pure decay case ( Fig-\ref{fig:allskymap}f),  substructures are less visible than in the pure annihilation scenario.  The smooth component 
also differs for the six scenarios in terms of both the central maximum and the profile slope, as shown in Fig-\ref{fig:fluxprof}. The adiabatic contraction case has a steep central profile such that the normalization of the total flux overwhelms the relatively small contribution from substructure. It looks more spherical in the all-sky map because of the assumption of spherical NFW$_c$ contraction (Fig-\ref{fig:allskymap}e). Both Sommefeld enhancement cases have a larger central flux than the pure annihilation case due to the distribution of the velocity dispersion.  With a $v^2$ dependence, the $p-$wave annihilation case therefore has a flatter central flux profile.
The pure decay case has the shallowest all-sky map and radial profile because of the linear $\rho$ dependence. 

We compare our results to the Fermi-LAT DGE
model. \citet{Ackermann:2012co} presented a measurement of the first
21 months of the Fermi-LAT mission and compared a grid of models of
DGE emission produced by the GALPROP code.  These models incorporate
the observed astrophysical distribution of cosmic-ray sources,
interstellar gas and radiation fields. They compare the predicted
model intensities and spectra with the observations in various sky
regions. The models are consistent with the Fermi data in the
anti-center regions but underpredict the data in the inner
 regions of the Galaxy at energies around a few GeV. 
 Their conclusions concerning the discrepancy mainly
focused on undetected point sources and cosmic ray spectral index
variations throughout the galaxy. Although it is generally accepted
that there is an ``excess" $\gamma$-ray signal most probably from the
 Galactic Center, it is still worth considering signals from
off-center regions. 

In \citet{Ackermann:2012co}, various sky regions are taken into
consideration for the comparison of models and observations.  
We take the same sky regions to plot the angular profiles, namely, 
$|b| > 8^\circ$, $|b| \le 5^\circ$, $|l|\le30^\circ$ and $|l|\ge90^\circ$. 
We also show the DGE model in each subplot. This DGE model is based on the GALPROP model
$^\mathrm{S}\mathrm{S}^\mathrm{Z}4^\mathrm{R}20^\mathrm{T}150^\mathrm{C}5$. 
Here we do not make comparisons to models with other choices of parameters.
However, since all the models are required to fit the
observational data, they show very similar features. In order to get the total gamma ray flux we
integrated over energy, from 50MeV to 800GeV, in the GALPROP
model.
The results are shown in Fig-\ref{fig:cutplots}. The upper left panel shows the longitudinal profile of the region of the sky with the central $|b|>8^\circ$ removed, while the upper right panel shows the longitudinal profile of the central $|b|<5^\circ$ of the Galaxy.
The lower left panel shows the latitude profiles
of the region with $|l|\le30^\circ$. 
The lower right panel shows the latitude profiles of the region with $|l|\le90^\circ$ removed. 
Two apparent large subhalos that are visible in the all-sky maps (Fig-\ref{fig:allskymap})  would make the latitude profiles off-central in this panel.
{
To better illustrate the comparison between the Fermi-LAT model and the profile from our model, we remove the influence from the 2 sub halos by  separating the smooth main halo component and the subhalo component. We select a random position sitting 8kpc away from the GC to get an all-sky map of the subhalo component such that there is no subhalo too bright or too close to the observer. After that, we combine the subhalo component and main halo component to get a properly centered profile}. 
The variations in the longitudinal and latitude profiles of the annihilation cases are the result of the contribution from the subhalos to the annihilation signal. The contribution from the subhalos is most important for the two Sommerfeld-corrected annihilation cases. The longitudinal profile of $1/v^2$ Sommerfeld-corrected case in the upper left panel is completely dominated by emission from subhalos.

{
Note that these profiles, if used to interpret the Fermi data, must be combined with the main DGE model, since they are in any case a small contribution to the full $\gamma$-ray emission. These comparisons provide us with intuition on how these different profiles would work if used to fit the real data. For example, in the upper panels, the DGE signal shows a relatively flat profile compared to the longitudinal profiles of the five annihilation cases. The decay profile is the most degenerate with the DGE signal and would provide a relatively larger weight when fit to data. On the contrary, in the lower two panels, due to the disk component of the Milky Way, the DGE signal has a  steep central profile that most closely resembles the latitude profile of the pure annihilation case. It may lead to a larger weight to the pure annihilation contribution. The subhalos' contribution cannot be directly used as a part of the profile model since their positions are arbitrary compared to the real case. As shown in the figures, this contribution is dominated by fluctuations and has  a flat envelope. They provide an additional flat smooth component if unresolved; and a measure of the statistical uncertainties, if resolved.}

\begin{figure}
    \centering
    \includegraphics[width=0.49\textwidth]{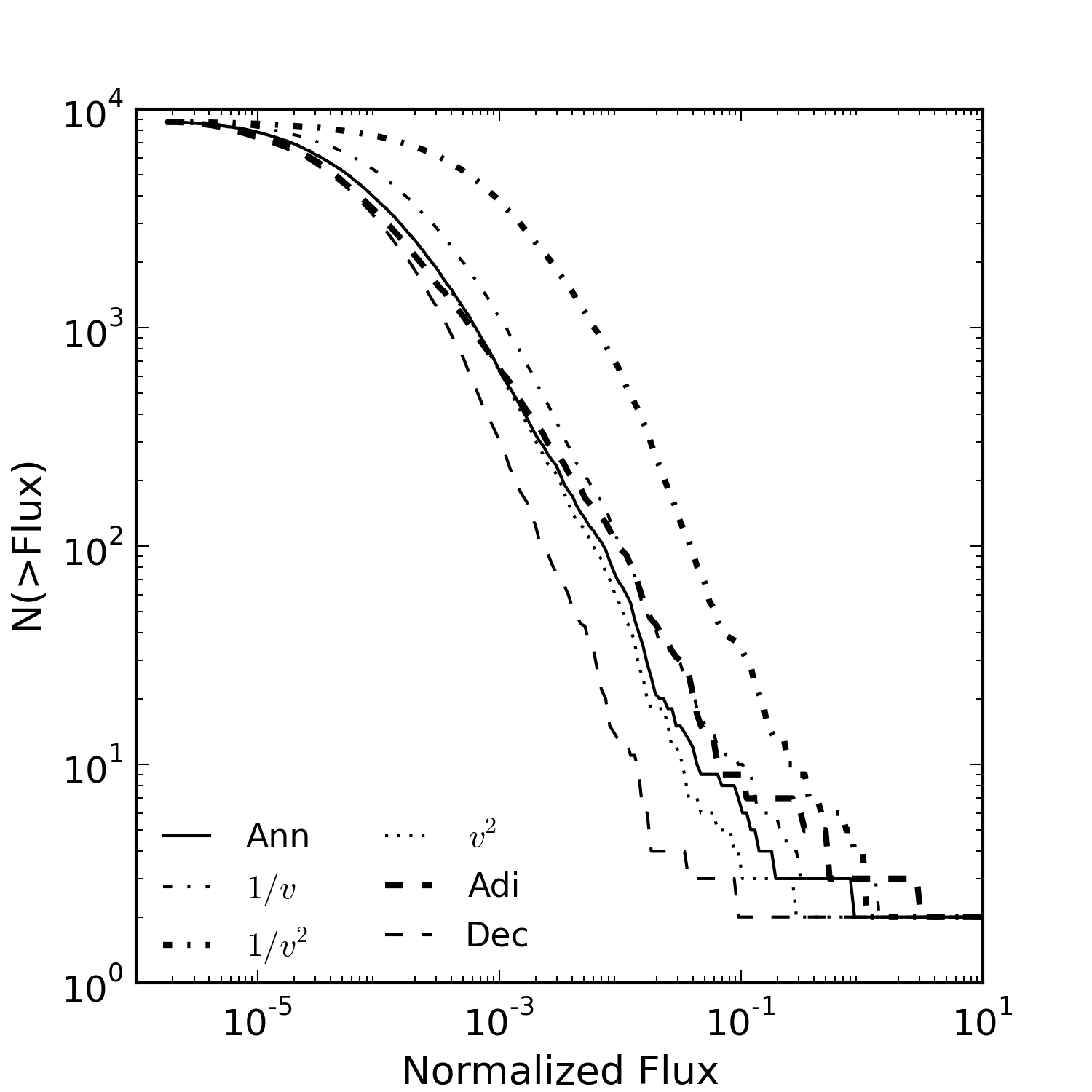}
    \caption{The flux function for different cases. {\it Solid}: pure
    annihilation. {\it Thin dash-dotted}: annihilation with Sommerfeld
    enhancement $1/v$ correction. {\it Thick dash-dotted}: annihilation with
    Sommerfeld enhancement $1/v^2$ correction.  {\it Dotted}: annihilation
    with $v^2$ correction. {\it Thick dashed}: annihilation with adiabatic
    contraction. {\it Thin dashed}: pure decay. The fluxes are normalized
    such that the host halo has flux unity. See text for details. }
    \label{fig:fluxfunc}
\end{figure}

We take the inner Galactic region    into special consideration since it
might be a promising region for detecting DM signals. Other
works propose alternative astrophysical models for the $\gamma$-ray
 excess at the Galactic Center, such as unresolved millisecond pulsars \citep{Abazajian:2011ea}. 
\citet{Hooper:2013ii} strongly disfavor this explanation for both the
spatial distribution and spectrum of observed millisecond pulsars. Assuming that
the $\gamma$-ray excess mainly comes from DM decays/annihilations, we
show here the predicted flux map (Fig-\ref{fig:center55}) and angular
profile (Fig-\ref{fig:fluxprof} right panel) of the inner
$5^\circ\times5^\circ$ region of the simulated maps for the different
$\gamma$-ray generating cases.  The angular profiles in
Fig-\ref{fig:fluxprof} right panel  are normalized to the total flux of
the host halo, as used in the all-sky maps.
The flux maps in Fig-\ref{fig:center55} are normalized such that the center pixel has
flux unity and the color scale is linear. 
The Sommerfeld-enhanced (Fig-\ref{fig:center55}b
and Fig-\ref{fig:center55}c) cases and adiabatic contraction case (Fig-\ref{fig:center55}e) have a 
more contracted profile than the pure annihilation case (Fig-\ref{fig:center55}a).  In contrast, the $p-$wave annihilation (Fig-\ref{fig:center55}d) and pure decay case (Fig-\ref{fig:center55}c) have flatter profiles illustrated by the greater spread in $\gamma$-ray flux over the $5^\circ\times5^\circ$ region. The six cases of the $\gamma$-ray flux maps of Fig-\ref{fig:center55} show considerable variations and complex asymmetrical contours in the central region. This asymmetry would be absent from simple parameterized model of the the DM annihilation signal from a NFW/cored profile considered previously \citep{Ackermann:2012ha}. {
Note that the simulation used in this paper has a convergence radius $380$ pc which corresponds to approximately $2.7^\circ$ in the center region. Therefore, our method would underestimate the real flux at the innermost region. To estimate this underestimation, we simply assume the real converged density profile is Equation-\ref{eqn:nfw} and the simulation profile is a cored profile $\rho(r) = \rho_s \exp\{-2/\alpha [(r/r_s)^\alpha-1]\}$ with $\alpha=0.170, r_s=21.5$kpc, $\rho_s=1.73\time10^{-3}\mathrm{M}_\sun$pc$^{-3}$\cite{Diemand:2008hr}. It turns out for annihilation, the NFW flux is 2 times larger of the cored profile occured at $0.8^\circ$; for decay, it occurred at $0.06^\circ$.  When using the templates to fit the real data, one should pay special attention the central few pixels.}

\begin{figure}
    \centering
    \includegraphics[width=0.49\textwidth]{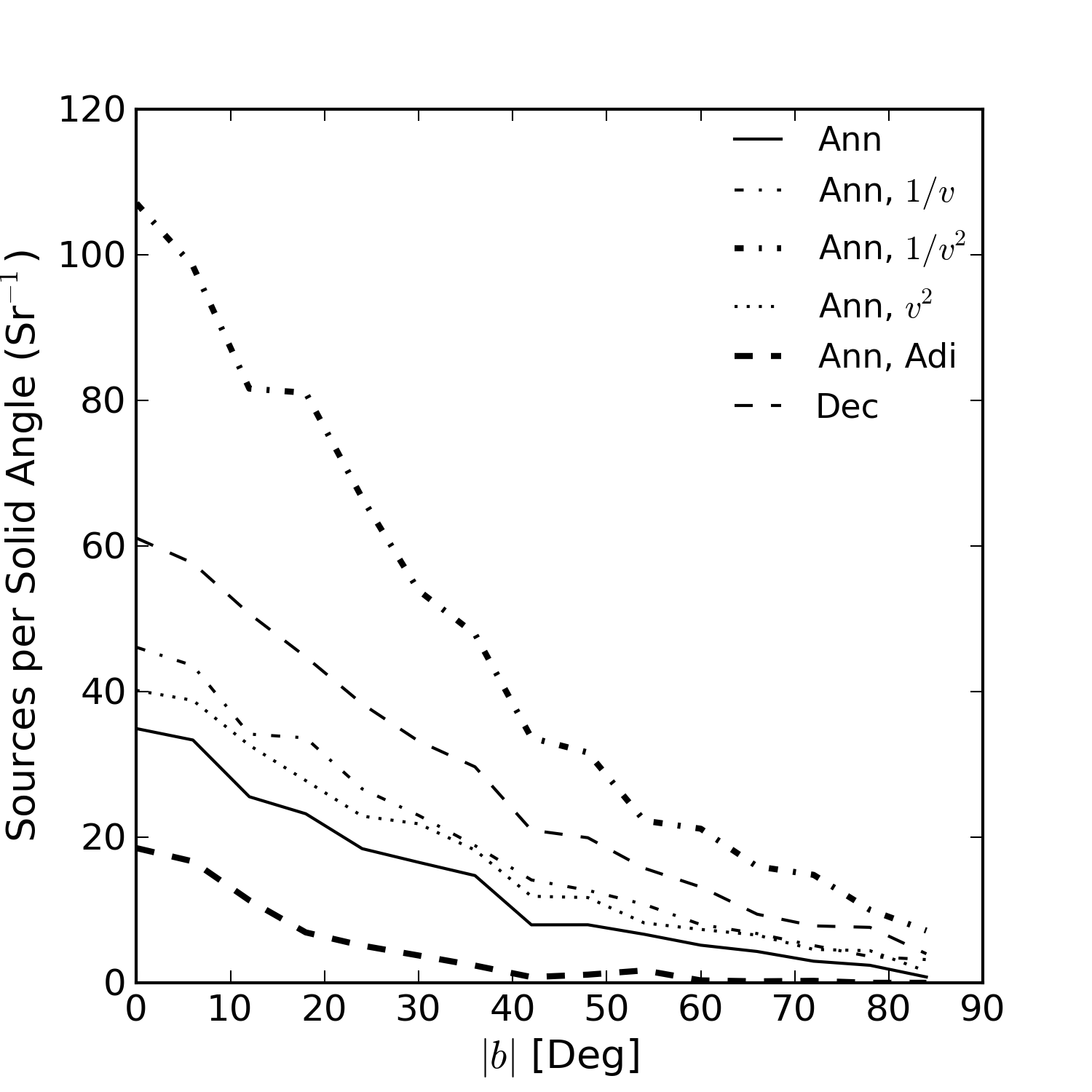}
    \caption{The latitude distribution of detected point sources. The
    detection threshold is chosen such that the maximum number of
    sources is on the order of 100. Sources are counted in latitude bins of 
    $10^\circ$. The legend is the same as 
    Fig. \ref{fig:fluxfunc}. See text for details.  }
    \label{fig:numfunc}
\end{figure}

\section{DISCUSSION\label{sec:discussion}}

{
The morphologies  and positions of each subhalo 
differ completely in different realizations 
of simulations, and may not directly apply to fitting 
templates to the observational data, whereas  their statistical properties can
be well studied using the maps produced by our tools. 
One useful statistical property is the angular power spectrum (APS) 
of anisotropies. As pointed out by \cite{Hensley:2013ig}, APS could be used along with 
the energy spectrum to decompose the emission components.  \cite{GomezVargas:2013ti} 
did actual comparisons using simulation-generated APS \cite{Fornasa:2013ig} with Fermi-LAT measurements \cite{Ackermann:2012ja}, 
yielding reasonable preliminary results that constrain the DM annihilation cross section. Our
results which consider six different physical scenarios present an  extension of these 
results. We leave a more detailed study of anisotropies (which must have the boost factor carefully 
calculated) to a future project. 
}

In addition to the diffuse emission, we are able to use our maps to
study the point sources from Fermi-LAT. For annihilations, when we add
in the Sommerfeld enhancement, the number density of detected small
halos increases. For the $1/v^2$ case, the substructure almost
dominates the radiation field other than in the galactic
center. Fig-\ref{fig:fluxfunc} shows the flux functions of sub-halos
for different cases. The flux of the subhalos are calculated by
summing over the flux of all pixels inside the halo's angular radius.
All the curves converge at $F\rightarrow 0$ since the total number of
sub-halos in the simulation is fixed at $\sim10^4$. The curve for the
Sommerfeld enhancement $1/v^2$ is much steeper than the other curves,
which indicates that the subhalos are brighter than in other
models. We further plot the detectable point sources (with angle
extending less than $\sim 2^\circ$ \citep{Belikov:2012gb}) in Fig-\ref{fig:numfunc}.
Compared to \citet{Belikov:2012gb}, 
these curves are  shallower than
the identified and unidentified sources in the second Fermi-LAT source
catalog for the inner region, but are  relatively flat at larger
latitudes. All of these features imply that observable DM signals could
come from off-center subhalos.

In summary, we are not aiming to set new constraints on DM
annihilations or decays, but rather to provide possible new templates 
{
(the smooth components)}
to fit the DGE data with dark matter models and illustrate the possible 
range of profiles for plausible models. 
Compared to a simple assumption that the DM halo has a NFW  or 
an isothermal profile, our results based on 
the simulation data are more realistic (e.g. Fig-\ref{fig:densityprofile}).
{
The effects of subhalos could be also be statistically included in the 
templates, e.g. APS or used to provide as  estimates of the 
uncertainties.
}
The Galactic central region might be a promising region for detecting a DM signal as recently pointed out by \citet{Gordon:2013jj}. 
The various DM maps we present here show
significant observational differences when the DM
annihilation/decay physics are changed.  
The morphology of the center 
region is also asymmetric and complex, 
and simple analytical templates may fail
to match the real signal.
There are three key points that we wish to
emphasize here.

Firstly, compared to the Fermi DGE result, the annihilation signal,
with or without Sommerfeld enhancement, has distinctly different central profiles. 
The shape of the Fermi excess profile in the Fermi bubble region has recently been used as a means
of discriminating dark matter from millisecond pulsar models of the
central excess \citep{Hooper:2013tl}.  Our philosophy here is similar
but we extend this approach to probing the nature of DM
annihilations. We predict different profiles for the alternative dark
matter annihilation models. The differences are especially notable for decaying dark matter, for which
the flux profile is very flat.

Secondly, it is clear that some of our models, especially the
Sommerfeld cases, provide more bright substructures than are allowed by the
observations in terms of unresolved and unidentified point sources. We
leave it to others to make detailed comparisons with the data, but our
predicted cumulative counts of subhalos should provide a basic
{comparison
}. The $p-$wave annihilation case is a promising model since it
removes many of these bright substructures. The model including adiabatic contraction has a similar effect of de-emphasizing the substructure
by increasing the contrast between host halo core and the
(unadiabatically contracted) subhalos, although the quantitative role of
adiabatic contraction of the DM in halos remains
controversial.
 
 Finally we note that we have not included any spectral signatures,
 which differ among the various adopted annihilation channels and particle masses: for
 example the Sommerfeld models require high values for the particle mass ($\sim$ TeV)
and hence would be best constrained through observational data at relatively high photon energies.
\begin{acknowledgments}
We thank Michael Kuhlen for his help on developing our algorithm. We acknowledge support from National Science Foundation grant OIA-1124403, and OCI-1040114.
The research of JS has also been supported at IAP by  ERC project 267117 (DARK) hosted by Universit\'e Pierre et Marie Curie - Paris 6.  Support for P.M. was provided by the NSF through grants OIA-1124453, 
and by NASA through grant NNX12AF87G. The research of A.S. and of RFGW was also supported by grant 109285 from the Gordon and Betty Moore Foundation.
\end{acknowledgments}

\end{document}